\documentclass[3p,times,10pt]{elsarticle}
\usepackage[utf8]{inputenc}
\usepackage[colorlinks = true,
            linkcolor = blue,
            urlcolor  = blue,
            citecolor = blue,
            anchorcolor = blue]{hyperref}
\usepackage{cleveref}
\usepackage{graphicx}
\usepackage{wrapfig}
\usepackage{float}
\usepackage[tablename=TABLE, labelsep=period, figurename=FIG., skip=0pt, font=small]{caption}
\usepackage{pdfpages}
\date{} 
\makeatletter
\def\ps@pprintTitle{%
 \let\@oddhead\@empty
 \let\@evenhead\@empty
 \def\@oddfoot{\centerline{\thepage}}%
 \let\@evenfoot\@oddfoot}
\makeatother
\def\Journal#1#2#3#4{{#1} {\bf #2}, #3 (#4)}

\def\NIMA{{\em Nucl. Instrum. Methods} A}

\def\be{\begin{equation}}
\def\ee{\end{equation}}
\def\bea{\begin{eqnarray}}
\def\eea{\end{eqnarray}}

\begin{document}
\begin{frontmatter}
\title{\textbf{The Low Polonium Field of Borexino and its significance for the CNO neutrino detection}}

\author[Juelich,RWTH]{\textbf{Sindhujha Kumaran}}
\author[]{\textbf{for the Borexino collaboration\\}}
\author[London1,Munchen]{M.~Agostini}
\author[Munchen]{K.~Altenm\"{u}ller}
\author[Munchen]{S.~Appel}
\author[Kurchatov]{V.~Atroshchenko}
\author[Juelich]{Z.~Bagdasarian\footnote{Present address: University of California, Berkeley, Department of Physics, CA 94720, Berkeley, USA}}
\author[Milano]{D.~Basilico}
\author[Milano]{G.~Bellini}
\author[PrincetonChemEng]{J.~Benziger}
\author[LNGS]{R.~Biondi}
\author[Milano]{D.~Bravo\footnote{Present address: Universidad Autónoma de Madrid, Ciudad Universitaria de Cantoblanco, 28049 Madrid, Spain}}
\author[Milano]{B.~Caccianiga}
\author[Princeton]{F.~Calaprice}
\author[Genova]{A.~Caminata}
\author[Virginia]{P.~Cavalcante\footnote{Present address: INFN Laboratori Nazionali del Gran Sasso, 67010 Assergi (AQ), Italy}}
\author[Lomonosov]{A.~Chepurnov}
\author[Milano]{D.~D'Angelo}
\author[Genova]{S.~Davini}
\author[Peters]{A.~Derbin}
\author[LNGS]{A.~Di Giacinto}
\author[LNGS]{V.~Di Marcello}
\author[Princeton]{X.F.~Ding}
\author[Princeton]{A.~Di Ludovico} 
\author[Genova]{L.~Di Noto}
\author[Peters]{I.~Drachnev}
\author[Dubna,Milano]{A.~Formozov}
\author[APC]{D.~Franco}
\author[Princeton,GSSI]{C.~Galbiati}
\author[LNGS]{C.~Ghiano}
\author[Milano]{M.~Giammarchi}
\author[Princeton]{A.~Goretti\footnote{Present address: INFN Laboratori Nazionali del Gran Sasso, 67010 Assergi (AQ), Italy}}
\author[Juelich,RWTH]{A.S.~G\"ottel}
\author[Lomonosov,Dubna]{M.~Gromov}
\author[Mainz]{D.~Guffanti}
\author[LNGS]{Aldo~Ianni}
\author[Princeton]{Andrea~Ianni}
\author[Krakow]{A.~Jany}
\author[Munchen]{D.~Jeschke}
\author[Kiev]{V.~Kobychev}
\author[London,Atomki]{G.~Korga}
\author[LNGS]{M.~Laubenstein}
\author[Kurchatov,Kurchatovb]{E.~Litvinovich}
\author[Milano]{P.~Lombardi}
\author[Peters]{I.~Lomskaya}
\author[Juelich,RWTH]{L.~Ludhova}
\author[Kurchatov]{G.~Lukyanchenko}
\author[Kurchatov]{L.~Lukyanchenko}
\author[Kurchatov,Kurchatovb]{I.~Machulin}
\author[Mainz]{J.~Martyn}
\author[Milano]{E.~Meroni}
\author[Dresda]{M.~Meyer}
\author[Milano]{L.~Miramonti}
\author[Krakow]{M.~Misiaszek}
\author[Peters]{V.~Muratova}
\author[Munchen]{B.~Neumair}
\author[Mainz]{M.~Nieslony}
\author[Kurchatov,Kurchatovb]{R.~Nugmanov}
\author[Munchen]{L.~Oberauer}
\author[Mainz]{V.~Orekhov}
\author[Perugia]{F.~Ortica}
\author[Genova]{M.~Pallavicini}
\author[Munchen]{L.~Papp}
\author[Juelich,RWTH]{L.~Pelicci}
\author[Juelich,RWTH]{\"O.~Penek}
\author[Princeton]{L.~Pietrofaccia}
\author[Peters]{N.~Pilipenko}
\author[UMass]{A.~Pocar}
\author[Kurchatov]{G.~Raikov}
\author[LNGS]{M.T.~Ranalli}
\author[Milano]{G.~Ranucci}
\author[LNGS]{A.~Razeto}
\author[Milano]{A.~Re}
\author[Juelich,RWTH]{M.~Redchuk\footnote{Present address: Dipartimento di Fisica e Astronomia dell’Università di Padova and INFN Sezione di Padova, Padova, Italy}}
\author[Perugia]{A.~Romani}
\author[LNGS]{N.~Rossi}
\author[Munchen]{S.~Sch\"onert}
\author[Peters]{D.~Semenov}
\author[Juelich]{G.~Settanta}
\author[Kurchatov,Kurchatovb]{M.~Skorokhvatov}
\author[Juelich,RWTH]{A.~Singhal}
\author[Dubna]{O.~Smirnov}
\author[Dubna]{A.~Sotnikov}
\author[LNGS,Kurchatov]{Y.~Suvorov\footnote{Present address: Dipartimento di Fisica, Universit\`a degli Studi Federico II e INFN, 80126 Napoli, Italy}}
\author[LNGS]{R.~Tartaglia}
\author[Genova]{G.~Testera}
\author[Dresda]{J.~Thurn}
\author[Peters]{E.~Unzhakov}
\author[LNGS,Aquila]{F.L.~Villante}
\author[Dubna]{A.~Vishneva}
\author[Virginia]{R.B.~Vogelaar}
\author[Munchen]{F.~von~Feilitzsch}
\author[Krakow]{M.~Wojcik}
\author[Mainz]{M.~Wurm}
\author[Genova]{S.~Zavatarelli}
\author[Dresda]{K.~Zuber}
\author[Krakow]{G.~Zuzel}

\address[APC]{AstroParticule et Cosmologie, Universit\'e Paris Diderot, CNRS/IN2P3, CEA/IRFU, Observatoire de Paris, Sorbonne Paris Cit\'e, 75205 Paris Cedex 13, France}
\address[Dubna]{Joint Institute for Nuclear Research, 141980 Dubna, Russia}
\address[Genova]{Dipartimento di Fisica, Universit\`a degli Studi e INFN, 16146 Genova, Italy}
\address[Krakow]{M.~Smoluchowski Institute of Physics, Jagiellonian University, 30348 Krakow, Poland}
\address[Kiev]{Kiev Institute for Nuclear Research, 03680 Kiev, Ukraine}
\address[Kurchatov]{National Research Centre Kurchatov Institute, 123182 Moscow, Russia}
\address[Kurchatovb]{ National Research Nuclear University MEPhI (Moscow Engineering Physics Institute), 115409 Moscow, Russia}
\address[LNGS]{INFN Laboratori Nazionali del Gran Sasso, 67010 Assergi (AQ), Italy}
\address[Milano]{Dipartimento di Fisica, Universit\`a degli Studi e INFN, 20133 Milano, Italy}
\address[Perugia]{Dipartimento di Chimica, Biologia e Biotecnologie, Universit\`a degli Studi e INFN, 06123 Perugia, Italy}
\address[Peters]{St. Petersburg Nuclear Physics Institute NRC Kurchatov Institute, 188350 Gatchina, Russia}
\address[Princeton]{Physics Department, Princeton University, Princeton, NJ 08544, USA}
\address[PrincetonChemEng]{Chemical Engineering Department, Princeton University, Princeton, NJ 08544, USA}
\address[UMass]{Amherst Center for Fundamental Interactions and Physics Department, University of Massachusetts, Amherst, MA 01003, USA}
\address[Virginia]{Physics Department, Virginia Polytechnic Institute and State University, Blacksburg, VA 24061, USA}
\address[Munchen]{Physik-Department, Technische Universit\"at  M\"unchen, 85748 Garching, Germany}
\address[Lomonosov]{Lomonosov Moscow State University Skobeltsyn Institute of Nuclear Physics, 119234 Moscow, Russia}
\address[GSSI]{Gran Sasso Science Institute, 67100 L'Aquila, Italy}
\address[Dresda]{Department of Physics, Technische Universit\"at Dresden, 01062 Dresden, Germany}
\address[Mainz]{Institute of Physics and Excellence Cluster PRISMA+, Johannes Gutenberg-Universit\"at Mainz, 55099 Mainz, Germany}
\address[Juelich]{Institut f\"ur Kernphysik, Forschungszentrum J\"ulich, 52425 J\"ulich, Germany}
\address[RWTH]{RWTH Aachen University, 52062 Aachen, Germany}
\address[Aquila]{Dipartimento di Scienze Fisiche e Chimiche, Universit\`a dell'Aquila, 67100 L'Aquila, Italy}
\address[London]{Department of Physics, Royal Holloway, University of London, Department of Physics, School of Engineering, Physical and
Mathematical Sciences, Egham, Surrey, TW20 OEX, UK}
\address[London1]{Department of Physics and Astronomy, University College London, London, UK}
\address[Atomki]{Institute of Nuclear Research (Atomki), Debrecen, Hungary}

\begin{abstract}
    
Borexino is a liquid scintillator detector located at the Laboratori Nazionale del Gran Sasso, Italy with the main goal to measure solar neutrinos. The experiment recently provided the first direct experimental evidence of CNO-cycle neutrinos in the Sun, rejecting the no-CNO signal hypothesis with a significance greater than 5$\sigma$ at 99\%C.L. The intrinsic $^{210}$Bi is an important background for this analysis due to its similar spectral shape to that of CNO neutrinos. $^{210}$Bi can be measured through its daughter $^{210}$Po which can be distinguished through an event-by-event basis via pulse shape discrimination. However, this required reducing the convective motions in the scintillator that brought additional $^{210}$Po from peripheral sources. This was made possible through the thermal insulation and stabilization campaign performed between 2015 and 2016. This article will explain the strategy and the different methods performed to extract the $^{210}$Bi upper limit in Phase-III (Jul 2016- Feb 2020) of the experiment through the analysis of $^{210}$Po in the cleanest region of the detector called the Low Polonium Field.
\end{abstract}

\end{frontmatter}
\section{CNO neutrino detection with Borexino}\label{sec:cno}
Solar neutrinos are elementary particles that are produced inside the Sun, by the same nuclear fusion processes that generate the heat. Neutrinos interact rarely after their production and are therefore a direct probe of these nuclear fusion processes. According to the Standard Solar Model (SSM), which represents the best knowledge available about the Sun, the heat in the Sun's core is generated by two main series of processes, fusing protons to Helium: the primary proton--proton ($pp$) chain, responsible for about 99\% of the solar energy production and the sub-dominant Carbon – Nitrogen – Oxygen (CNO) cycle. Neutrinos are detected via their elastic scattering on electrons in the liquid scintillator. Borexino~\cite{bx} has already published a complete spectroscopy of $pp$ chain neutrinos~\cite{pp} and has recently provided the first direct experimental evidence of solar neutrinos produced in the rare CNO nuclear fusion cycle~\cite{cno}. The main challenges of this analysis are the very low interaction rate of CNO neutrinos and the similarity of its spectral shape to that of $pep$ solar neutrinos and the intrinsic $^{210}$Bi background. The $pep$ neutrino rate can be independently determined with 1.4\% precision using the constraint on solar luminosity, global analysis with all solar neutrino experiments excluding the latest Borexino data, exploiting theoretically precisely known ratio of $pep$ and $pp$ neutrino fluxes, and using the most recent values of the oscillation parameters. The $^{210}$Bi background\textemdash the short-living decay product of $^{210}$Pb, can be determined via the counting of $\alpha$-decays of its daughter $^{210}$Po. This assumes secular equilibrium of the chain down from $^{210}$Pb, which is a long-living isotope contaminating the liquid scintillator. Alpha particles can be identified in Borexino through pulse shape discrimination techniques. Until mid-2016, additional $^{210}$Po was brought from peripheral sources to the fiducial volume through the convective motions of the scintillator, triggered by seasonal temperature changes. However, between 2015 and 2016, the detector was thermally insulated and an active temperature control system was installed. This has minimized the residual convection in the innermost parts of the detector, making it possible to measure $^{210}$Bi via $^{210}$Po and has helped in obtaining a $^{210}$Bi upper limit from the cleanest region of the detector called the Low Polonium Field. A multivariate fit was then performed, i.e. the
energy spectra in the window between 320\,keV and 2,640\,keV and the
radial distribution of the Phase-III data (July 2016 - February 2020) was simultaneously fitted, after constraining the rates of $pep$ and $^{210}$Bi. This study excluded the no-CNO signal scenario with a significance greater than 5.0$\sigma$ at 99.0\% CL. The total contribution of the systematics was evaluated as $_{-0.5}^{+0.6}$\,cpd/100t using 13.8 million pseudo-datasets with the same exposure as Phase-III. The systematic uncertainties included the $^{210}$Bi spectral shape, the energy scale and resolution of the Monte Carlo model, non-linearity and non-uniformity of the detector's response, as well as variation in the absolute value of the scintillator light yield. A simple counting analysis, complementary to the multivariate fit, rejected the null CNO hypothesis at 3.5$\sigma$. 

\section{The Low Polonium Field}\label{sec:lpof}
The Low Polonium Field of Borexino developed just above the equator of the detector around mid-2016. The $^{210}$Po rate in this field is the sum of two components: a {\it scintillator} component that comes from the $^{210}$Pb in the scintillator and is assumed to be in secular equilibrium with $^{210}$Bi, and a {\it vessel} component which is due to the migration of $^{210}$Po from the $^{210}$Pb on the vessel. The migration process is driven by convective currents. However, the parent $^{210}$Pb and $^{210}$Bi isotopes of this component stay on the IV. The qualitative shape and approximate position of the LPoF have been reproduced by fluid dynamical numerical simulations~\cite{cfd}. The $^{210}$Po content is not spatially uniform in the LPoF, but, the presence of a clear and consistent minimum helps in obtaining a robust $^{210}$Bi upper limit. The center of the LPoF is very stable over the entire Phase-III and slowly moves only by less than 20\,cm per month. This can be observed in Figure~\ref{fig:LPoF}(left), where the $^{210}$Po rate evolution over time in Phase-III is plotted for $z$-slices of height 0.1\,m with a 3\,m radial cut applied along the $\rho$ plane. The LPoF in the detector resembles an ellipsoid of $\sim$20 tons with rotational symmetry along the $x-y$ plane and with a bit of complexity along the $z$-axis.

\section{Analysis strategy and methods}\label{sec:fit}
The $^{210}$Po events in the LPoF are chosen using a geometrically normalised charge variable in the range 150-270\,p.e ($\sim$300-540\,keV). In addition, a pulse shape discrimination cut using the Multi-Layer Perceptron (MLP) developed via deep-learning techniques is applied such that only events below an MLP value of 0.3 (i.e. $\alpha$-like) are chosen. The paraboloid equation in 2D assuming rotational symmetry along the $x-y$ plane that is used to fit the data in order to obtain the minimum $^{210}$Po rate ($R(^{210}Po_{min})$) is as follows:
\begin{equation}\label{eq:bi_fit}
\frac{d^{2}R_{Po}}{d(\rho^{2})dz} = [R(^{210}Po_{min})\epsilon_{E}\epsilon_{MLP} + R_{\beta}]\bigg(1 + \frac{\rho^{2}}{a^{2}} + \frac{(z - z_{0})^{2}}{b^{2}} \bigg). 
\end{equation}
Here $\rho^{2} = x^{2} + y^{2}$, $z_{0}$ is the minimum position of the LPoF along the $z$-axis, $a$ and $b$ are shape parameters along the respective axes. $\epsilon_{E}$ and $\epsilon_{MLP}$ are the efficiency of the energy and pulse shape discrimination cuts, respectively. The efficiencies can be calculated using using a pure MC sample of $^{210}$Po events. $R_{\beta}$ is the rate of $\beta$ events in the region where $^{210}$Po events are selected. $R_{\beta}$ is obtained in an independent way through an energy fit performed using the selected $^{210}$Po events, the MC PDF of $^{210}$Po, and the MC PDFs of all the $\beta$-components, namely, $^{7}$Be, CNO, {\it pep}, $^{210}$Bi, {\it pp}, $^{14}$C, and $^{85}$Kr.\\
    \begin{figure}[t]
\centering
    \includegraphics[width=0.47\textwidth]{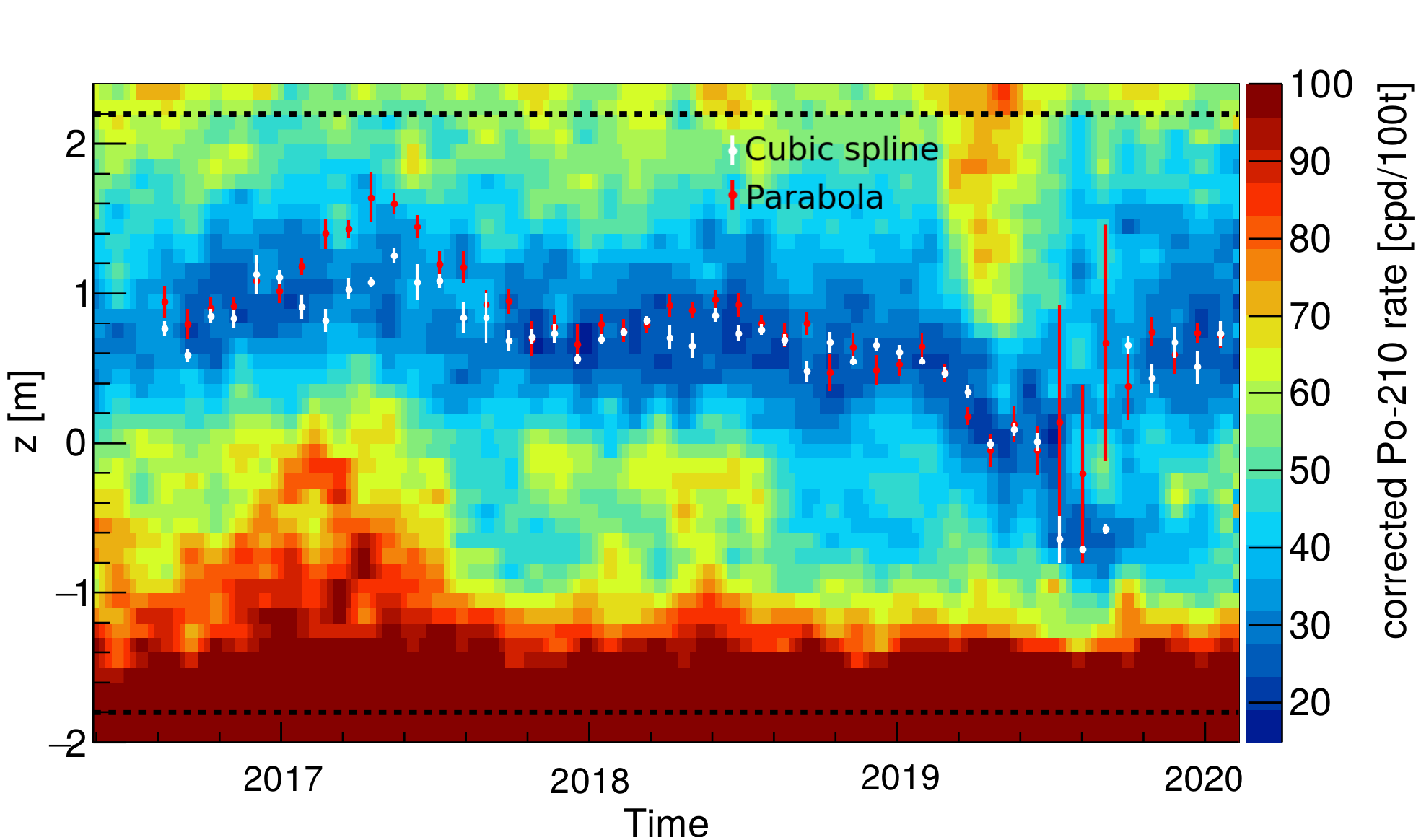}
    \includegraphics[width=0.47\textwidth]{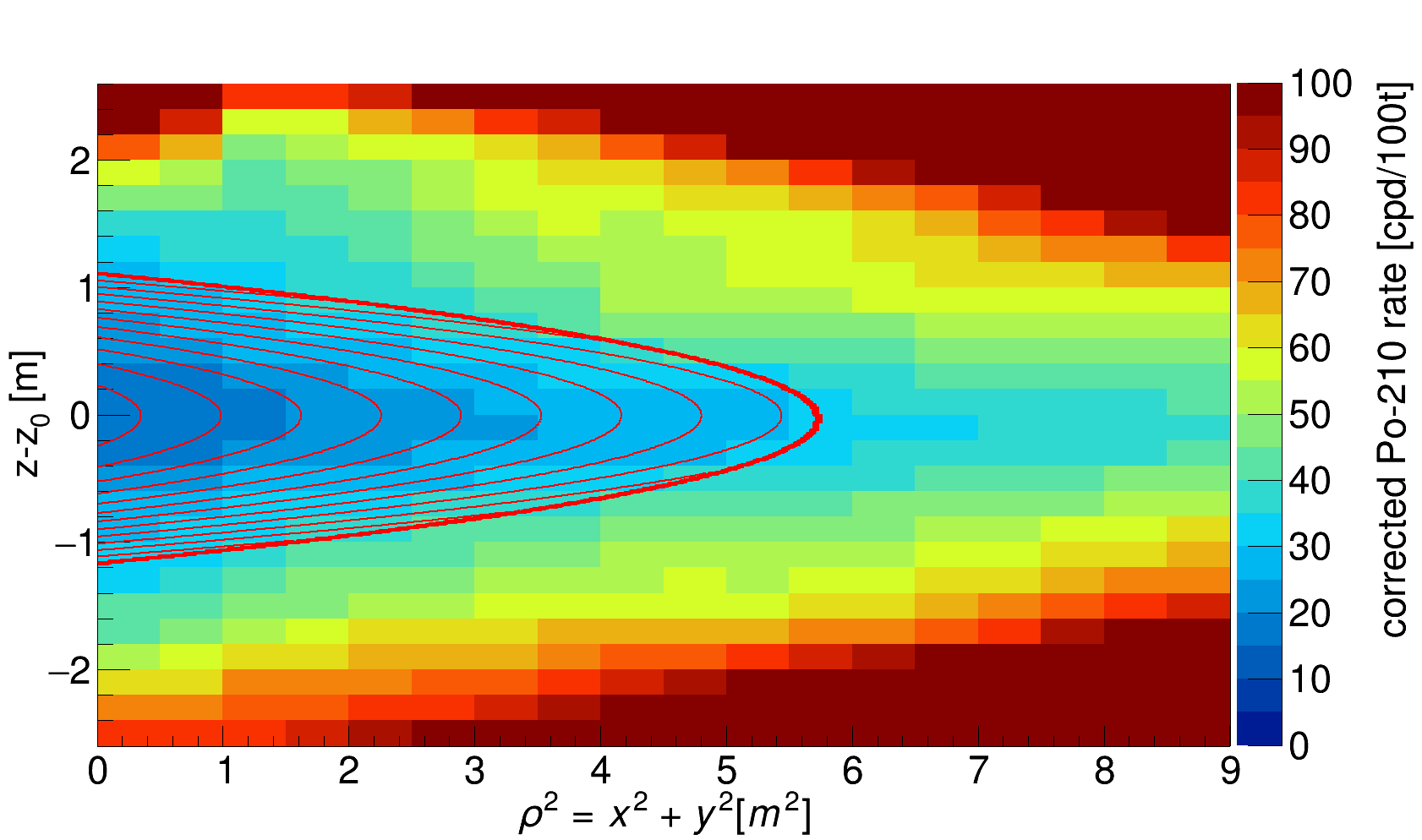}
   \caption{Left: Evolution of the LPoF over time in Phase-III along the $z$-axis, with the centers obtained from different methods. The radius considered in the $\rho$ plane is 3\,m. The black dashed lines represent the fiducial volume of the CNO analysis. Right: $\rho^{2}-z$ projection of the aligned LPoF fitted with equation~\ref{eq:bi_fit}. The fit function is shown in red. The colour scheme in both figures represent the $^{210}$Po rate in cpd/100t after correcting for the $\alpha$-detection inefficiency and $\beta$-leakage. Figures taken from~\cite{cno}.}
   \label{fig:LPoF}
 \end{figure}
Since the LPoF slowly moves along the $z$-axis due to residual convective motions, the data in the LPoF needs to be aligned along the $z$-direction before performing the fit on the full dataset. This is done by ``blind'' alignment of the data in the LPoF every month (or every two months) using the center $z_{0}$ obtained by fitting the data of the previous month. These monthly fits are performed in bigger volumes of 70 tons or 100 tons owing to the low monthly statistics in 20 tons. After the blind alignment using the centers of every month, the final fit is performed on the aligned dataset in around 20 tons ($\sim$5000 events) using either a simple paraboloid in equation~\ref{eq:bi_fit} with four free parameters ($R(^{210}Po_{min}), a, b, z_{0}$) or with more free parameters as it will described later in this section. The simple paraboloid fit can be performed either as a likelihood fit with ROOT (v. 6.22/03)~\cite{root} or with a Bayesian tool called MultiNest~\cite{MN}. The final aligned dataset with a simple paraboloid fit performed with ROOT is shown in Figure~\ref{fig:LPoF}(right).

MultiNest is a widely-used importance nested sampling algorithm.``Nested Sampling'' is a Monte Carlo technique used to estimate the Bayesian evidence, and generates posterior samples as a by-product. In addition to the 2D paraboloid fit, 3D ellipsoidal fits were also performed with MultiNest without assuming rotational symmetry along the $x-y$ plane, which resulted in statistically compatible results. The assumption of the rotational symmetry was also verified based on the fit results. Since the LPoF data is complex along the $z$-direction, especially during some thermal operations performed in summer of 2019, the 2D paraboloid fit was also performed with the implementation of a cubic spline along the $z$-axis in MultiNest. A cubic spline is a piece-wise defined polynomial of third degree. It is expressed through so-called knots. The space between two knots is filled with a polynomial. The complexity of the cubic spline for the LPoF was chosen based on the Bayesian evidence of the fit and it varied between 4 and 11 knots over the entire Phase-III. The average complexity for the 70/100-ton monthly fits was 7 knots. The final aligned fit in 20 tons was performed with 5 knots. Even though this method provided the best Bayesian evidence when compared to the simple paraboloid, the results were statistically compatible. The final methods used for the estimation of the $^{210}$Bi upper limit were a simple 2D paraboloid fit and a 2D paraboloid fit with cubic spline along the $z$-axis, performed with MultiNest.

In order to confirm that there is no bias in the $^{210}$Bi upper limit calculated and to check the compatibility between different fit methods, a toy MC validation was performed. 140 toy MC datasets, with realistic and extreme scenarios, with livetime of 2 years each, were generated with two components, namely,  {\it scintillator} $^{210}$Po with uniform distribution, and {\it vessel} $^{210}$Po with exponential weight, evaluated based on the distance from the inner vessel. It was found that the different methods were compatible with each other for different combinations of both the $^{210}$Po components. The method of using the Bayesian evidence to choose the complexity of the spline was also verified, i.e. due to the perfect ellipsoidal shape of the toy MC datasets, the best evidence was always obtained for the simple paraboloid fit and the cubic spline fit with the least complexity (4 knots). The extracted $R(^{210}Po_{min})$ from 500 datasets with a realistic scenario, was always positively biased for both the fit methods, proving that the $^{210}$Bi upper limit is conservative. Therefore, it does not result in false enhancement of the CNO neutrino rate.

\section{Sources of uncertainty}\label{sec:syst}
The different sources of uncertainty~\cite{neutel} that were considered for the $^{210}$Bi upper limit, in addition to the statistical uncertainty of the fit (0.83\,cpd/100t) are:
\begin{itemize}
    \item Systematic uncertainties from the fit: Mass of the fit region (0.40\,cpd/100t), and binning of the data histogram (0.20\,cpd/100t).
    \item Uncertainty on the $\beta$-leakage estimation, i.e. $R_{\beta}$ in equation~\ref{eq:bi_fit} (0.30\,cpd/100t).
    \item Homogeneity of $\beta$-events: Since the $^{210}$Bi upper limit is estimated from the LPoF which is only 20 tons, it is necessary to study the homogeneity of $\beta$-events in the entire fiducial volume of 70 tons, in the energy region of $^{210}$Bi. The radial homogeneity was studied by dividing the fiducial volume into 25 iso-volumetric shells (0.52\,cpd/100t). The angular homogeneity was studied by extending Fourier decomposition over a sphere surface, by projecting the spatial co-ordinates of the selected events on a sphere (0.58\,cpd/100t).
\end{itemize}

\section{Results}\label{sec:results}
The final $^{210}$Bi upper limit obtained through the estimation of the minimum $^{210}$Po rate in the Low Polonium Field of Borexino is 11.5 $\pm$ 1.3\,cpd/100\,t. This includes all the sources of uncertainty along with the statistical uncertainty of the fit. The best fit value for the CNO neutrino rate from the final multivariate spectral fit using this $^{210}$Bi upper limit is 7.2$^{+2.9}_{-1.7}$(stat)\,cpd/100\,t~\cite{cno}.

\end{document}